\documentclass{article}[9pt]
\usepackage{spconf,amsmath,epsfig}
\ninept

\title{Multi-Branch MMSE Decision Feedback Detection Algorithms
with Error Propagation Mitigation for Multi-Antenna Systems
 \vspace{-0.85em}} \twoauthors{ Rodrigo C. de Lamare \vspace{-1em}}{\small Communications Research Group,
  \\ \small University of York, United Kingdom \\
    \small Email: rcdl500@ohm.york.ac.uk \vspace{-1.75em}}{Didier Le
    Ruyet \vspace{-1em}}{\small Conservatoire National des Arts et Metiers, \\ \small Paris, France. \\ \small Email:
    leruyet@cnam.fr \vspace{-1.75em}}

\begin{document}
\maketitle

\begin{abstract}
In this work we propose novel decision feedback (DF) detection
algorithms with error propagation mitigation capabilities for
multi-input multi-output (MIMO) spatial multiplexing systems based
on multiple processing branches. The novel strategies for
detection exploit different patterns, orderings and constraints
for the design of the feedforward and feedback filters. We present
constrained minimum mean-squared error (MMSE) filters designed
with constraints on the shape and magnitude of the feedback
filters for the multi-branch MIMO receivers and show that the
proposed MMSE design does not require a significant additional
complexity over the single-branch MMSE design. The proposed
multi-branch MMSE DF detectors are compared with several existing
detectors and are shown to achieve a performance close to the
optimal maximum likelihood detector while requiring significantly
lower complexity. \vspace{-0.35em}

\end{abstract}

\begin{keywords}
{MIMO systems, spatial multiplexing, decision feedback detection,
constrained optimization.}
\end{keywords}
\vspace{-0.5em}

\section{Introduction}

Multiple transmit and receive antennas in wireless communication
systems can offer significant multiplexing \cite{foschini,vblast}
and diversity gains \cite{tarokh}. In a spatial multiplexing
configuration, the system can obtain substantial gains in data
rate. The capacity gains grow linearly with the minimum number of
transmit and receive antennas, and the transmission of individual
data streams from the transmitter to receiver \cite{foschini}. In
order to separate these streams, a designer must resort to MIMO
detection techniques. The optimal maximum likelihood (ML) detector
can be implemented using the sphere decoder (SD) algorithm
\cite{hassibi}. However, the computational complexity of this
algorithm depends on the noise variance, the number of data
streams to be detected and the signal constellation, resulting in
high costs for low signal-to-noise ratios (SNR), large MIMO
systems and large constellations. The complexity requirements of
the ML detector and the SD algorithm have motivated the
development of numerous alternative strategies for MIMO detection.
The linear detector \cite{duel_mimo}, the successive interference
cancellation (SIC) approach used in the VBLAST systems
\cite{vblast} and the decision feedback (DF) detectors
\cite{dhahir}-\cite{kusume} are techniques that can offer an
attractive trade-off between performance and complexity. An often
criticized aspect of these sub-optimal schemes is that they
typically do not achieve the diversity of the ML and SD
algorithms. This motivated the investigation of detectors such as
lattice-reduction-based schemes \cite{wuebben,gan} and calls for
new cost-effective algorithms with near-ML or ML performance.

In this work we propose novel DF detection strategies for MIMO
spatial multiplexing systems based on multiple processing branches
and error propagation mitigation. Prior work on DF schemes
includes the DF detector with SIC (S-DF) \cite{dhahir,kusume} and
the DF receiver with PIC (P-DF) \cite{woodward2}, combinations of
these schemes \cite{woodward2,delamare_mber,delamare_spadf} and
mechanisms to mitigate error propagation
\cite{reuter,delamare_itic}. The proposed detector employs
multiple feedforward and feedback filters and yields multiple
decision candidates. The proposed structure exploits different
patterns and orderings and selects the candidate and branch which
yield the estimates with the highest likelihood. We present
constrained minimum mean-squared error (MMSE) filters designed
with constraints on the shape and the magnitude of the feedback
filters for the multi-branch MIMO receivers. We show that the
proposed MMSE design does not require a significant additional
complexity since it relies on similar filter realizations but with
different constraints on the feedback filters. The proposed
multi-branch MMSE DF detectors are compared with several existing
detectors via simulations. The main contributions of this work are as follows: \\
1) Multi-branch MMSE decision feedback detectors ;
\\ 2) MMSE filter expressions with shape and magnitude constraints; \\
3) A study of the proposed and existing MIMO detectors.


\section{System Model}

Let us consider a spatial multiplexing MIMO system with $N_T$
transmit antennas and $N_R$ receive antennas, where $N_R \geq
N_T$. At each time instant $[i]$, the system transmits $N_T$
symbols which are organized into a $N_T \times 1$ vector
${\boldsymbol s}[i] = \big[ s_1[i], ~s_2[i], ~ \ldots,~ s_{N_T}[i]
\big]^T$ taken from a modulation constellation $A = \{
a_1,~a_2,~\ldots,~a_N \}$, where $(\cdot)^T$ denotes transpose.
The symbol vector ${\boldsymbol s}[i]$ is then transmitted over
flat fading channels and the signals are demodulated and sampled
at the receiver, which is equipped with $N_R$ antennas. The
received signal after demodulation, matched filtering and sampling
is collected in an $N_R \times 1$ vector ${\boldsymbol r}[i] =
\big[ r_1[i], ~r_2[i], ~ \ldots,~ r_{N_R}[i] \big]^T$ with
sufficient statistics for detection and given by
\begin{equation}
{\boldsymbol r}[i] = {\boldsymbol H} {\boldsymbol s}[i] +
{\boldsymbol n}[i],
\end{equation}
where the $N_R \times 1$ vector ${\boldsymbol n}[i]$ is a zero
mean complex circular symmetric Gaussian noise with covariance
matrix $E\big[ {\boldsymbol n}[i] {\boldsymbol n}^H[i] \big] =
\sigma_n^2 {\boldsymbol I}$, where $E[ \cdot]$ stands for expected
value, $(\cdot)^H$ denotes the Hermitian operator, $\sigma_n^2$ is
the noise variance and ${\boldsymbol I}$ is the identity matrix.
The symbol vector ${\boldsymbol s}[i]$ has zero mean and a
covariance matrix $E\big[ {\boldsymbol s}[i] {\boldsymbol s}^H[i]
\big] = \sigma_s^2 {\boldsymbol I}$, where $\sigma_s^2$ is the
signal power. The elements $h_{n_R,n_T}$ of the $N_R \times N_T$
channel matrix ${\boldsymbol H}$ are the complex channel gains
from the $n_T$th transmit antenna to the $n_R$th receive antenna.

\section{Multi-Branch MMSE  DF Detectors}

In this section, we detail the proposed multi-branch MMSE decision
feedback (MB-MMSE-DF) structure for MIMO systems. The proposed
MB-MMSE-DF scheme,  shown in Fig. 1, employs multiple signal
processing branches with appropriate shape constraints that modify
the design of both feedforward and feedback filters. The detector
exploits different patterns and orderings for the design of the
feedforward and feedback filters. The aim is to mitigate the error
propagation and approach the performance of the optimal ML
detector. The proposed MB-MMSE-DF scheme can achieve the full
diversity available in the system by increasing the number of
branches and, therefore, the number of candidate symbols prior to
detection.

\vspace*{-1.0em}
\begin{figure}[htb]
       \centering  
       \hspace*{-3.5em}
        \vspace*{-3.25em}
      {\includegraphics[width=10cm, height=5.0cm]{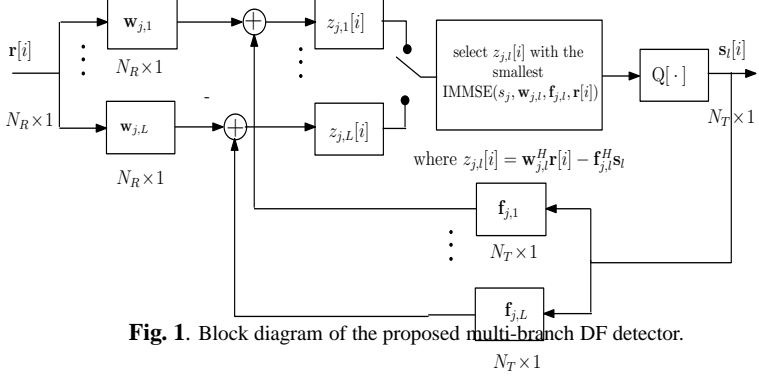}}
       \vspace*{-1.5em}
       \caption{\footnotesize Block diagram of the proposed multi-branch DF
detector.} \label{fig:mbdf}
\end{figure}

The MB-MMSE-DF employs multiple feedforward and feedback filters
such that it can obtain different local maxima of the likelihood
function associated with the ML detector. In order to detect each
transmitted data stream, the proposed MB-MMSE-DF detector linearly
combines the feedforward filter ${\boldsymbol w}_{j,l}$
corresponding to the $j$-th data stream and the $l$-th branch with
the received vector ${\boldsymbol r}[i]$, subtracts the remaining
interference by linearly combining the feedback filter
${\boldsymbol f}_{j,l}$ with the $N_T \times 1$ vector of initial
decisions $\hat{\boldsymbol s}_{j,l}^o[i]$ obtained from previous
decisions. This input-output relation is given by
\begin{equation}
\begin{split}
\label{eq:df} { z}_{j,l}[i] & = {\boldsymbol
w}^{H}_{j,l}{\boldsymbol r}[i] - {\boldsymbol f}_{j,l}^{H}
\hat{\boldsymbol s}_{j,l}^o[i], \\ j & = 1, ~ \ldots, ~N_T~~ {\rm
and} ~~ l=1,~\ldots,~L,
\end{split}
\end{equation}
where the input to the decision device for the $i$th symbol and
$j$-th stream is the $L \times 1$ vector ${\boldsymbol z}_j[i]
=[z_{j,1}[i] ~\ldots ~z_{j,L}[i]]^{T}$. The number of parallel
branches $L$ that yield detection candidates is a parameter that
must be chosen by the designer. In this context, the optimal
ordering algorithm conducts an exhaustive search $L=N_T!+1$ and is
clearly very complex for practical systems when $N_T$ is
significantly large. Our goal is to employ a reduced number of
branches and yet achieve near-ML or ML performance. The proposed
MB-MMSE-DF detector selects the best branch according to
\begin{equation}
l_{{\rm opt}} = \arg \min_{1 \leq l \leq L}   {\rm iMMSE}
(s_j[i],{\boldsymbol w}_{j,l}, {\boldsymbol f}_{j,l}), ~j=1, ~\ldots
,~ N_T \label{eq:error}
\end{equation}
where ${\rm iMMSE} (s_j[i],{\boldsymbol w}_{j,l}, {\boldsymbol
f}_{j,l})$ corresponds to the instantaneous MMSE produced by the
pair of filters ${\boldsymbol w}_{j,l}$ and ${\boldsymbol f}_{j,l}$.
The final detected symbol of the MB-MMSE-DF detector is obtained by
using the optimal branch:
\begin{equation}
\begin{split}
\hat{s}_{j}[i] & = Q \big[ {\boldsymbol z}_{j,l_{\rm opt}}[i]
\big]
\\ & = Q \big[ {\boldsymbol w}^{H}_{j,l_{\rm opt}}{\boldsymbol r}[i] -
{\boldsymbol f}_{j,l_{\rm opt}}^{H} \hat{\boldsymbol s}_{j,l_{\rm
opt}}^o[i]  \big], ~j=1, ~ \ldots,~ N_T \label{eq:dec}
\end{split}
\end{equation}
where $Q( \cdot)$ is a slicing function that makes the decisions
about the symbols, which is drawn from an M-PSK or a QAM
constellation. 

\subsection{MMSE Filter Design }

The design of the MMSE filters of the proposed MB-MMSE-DF detector
is equivalent to determining feedforward filters ${\boldsymbol
w}_{j,l}$ with dimensions $N_R \times 1$ and feedback filters
${\boldsymbol f}_{j,l}$ with $N_T \times 1$ elements subject to
certain shape and magnitude constraints on ${\boldsymbol f}_{j,l}$
in accordance with the following optimization problem
\begin{equation}
\begin{split}
\label{eq:msedfprop} {\rm min} & ~ {\rm MSE} (s_j[i],{\boldsymbol
w}_{j,l},{\boldsymbol f}_{j,l}) = E\big[ | { s}_j[i] -
{\boldsymbol w}^H_{j,l}{\boldsymbol r}[i] + {\boldsymbol
f}_{j,l}^{H}\hat{\boldsymbol s}_{j,l}^o[i] |^2 \big] \\
& {\rm subject}~{\rm to}~  {\boldsymbol S}_{j,l} {\boldsymbol
f}_{j,l} = {\boldsymbol v}_{j,l} ~~ {\rm and} ~~
 ||{\boldsymbol f}_{j,l}||^2 = \gamma_{j,l} || {\boldsymbol
f}_{j,l}^c ||^2 \\ {\rm for} ~ j & = 1, \ldots, N_T ~~{\rm and}~~
l = 1, \ldots, L,
\end{split}
\end{equation}
where the $N_T \times N_T$ shape constraint matrix is
${\boldsymbol S}_{j,l}$, ${\boldsymbol v}_{j,l}$ is the resulting
$ N_T \times 1$ constraint vector and ${\boldsymbol f}_{j,l}^c$ is
a feedback filter without constraints on the magnitude of its
squared norm.

By resorting to the method of Lagrange multipliers, computing the
gradient vectors of the Lagrangian with respect to ${\boldsymbol
w}_{j,l}$ and ${\boldsymbol f}_{j,l}$, equating them to null
vectors and rearranging the terms we obtain ${\rm for} ~ j  = 1,
\ldots, N_T$ and $l = 1, \ldots, L$
\begin{equation}
\label{eq:dfeprop1} {\boldsymbol w}_{j,l} = {\boldsymbol
R}^{-1}({\boldsymbol p}_{j} + {\boldsymbol Q}{\boldsymbol
f}_{j,l}),
\end{equation}
\begin{equation}
\begin{split}
\label{eq:dfeprop2} {\boldsymbol f}_{j,l} & = \beta_{j,l}
{\boldsymbol \Pi}_{j,l} ({\boldsymbol Q}^H{\boldsymbol w}_{j,l} -
{\boldsymbol t}_{j}) + ({\boldsymbol I} - {\boldsymbol \Pi}_{j,l})
{\boldsymbol v}_{j,l},
\end{split}
\end{equation}
where
\begin{equation}
{\boldsymbol \Pi}_{j,l} = {\boldsymbol I} - {\boldsymbol
S}_{j,l}^H({\boldsymbol S}_{j,l}^H {\boldsymbol
S}_{j,l})^{-1}{\boldsymbol S}_{j,l}
\end{equation}
is a projection matrix that ensures the shape constraint
${\boldsymbol S}_{j,l}$ and $\beta_{j,l} =  (1 -
\alpha_{j,l})^{-1}$ is a factor that adjusts the magnitude of the
feedback, $0 \leq \beta_{j,l} \leq 1$ and $\alpha_{j,l}$ is the
Lagrange multiplier. The relationship between $\beta_{j,l}$ and
$\gamma_{j,l}$ is not in closed-form even though we have
$\beta_{j,l} =0$ and $\beta_{j,l} =1 $ for $\gamma_{j,l}=0$
(standard linear MMSE detector) and $\gamma_{j,l}=1$ ( standard
MB-MMSE-DF detector), respectively. The $N_R \times N_R$
covariance matrix of the input data vector is ${\boldsymbol
R}=E[{\boldsymbol r}[i]{\boldsymbol r}^H[i]]$, ${\boldsymbol p}_j
= E[{\boldsymbol r}[i] s_j^*[i]]$, ${\boldsymbol Q} = E\big[
{\boldsymbol r}[i] \hat{\boldsymbol s}_{j,l}^{o,~H}[i] \big]$, and
${\boldsymbol t}_j = E[\hat{\boldsymbol s}_{j,l}^o[i] {s}_j^*[i]]$
is the $N_T \times 1$ vector of correlations between
$\hat{\boldsymbol s}_{j,l}^o[i]$ and ${s}_j^*[i]$.  Substituting
(\ref{eq:dfeprop2}) into (\ref{eq:dfeprop1}) and then further
manipulating the expressions for ${\boldsymbol v}_{j,l}=
{\boldsymbol 0}$ , we arrive at
\begin{equation}
\label{eq:dfeprope1} {\boldsymbol w}_{j,l} = \big( {\boldsymbol R}
- \beta_{j,l}{\boldsymbol Q} {\boldsymbol \Pi}_{j,l} {\boldsymbol
Q}^H \big)^{-1} \big({\boldsymbol p}_{j} - {\boldsymbol \Pi}_{j,l}
{\boldsymbol t}_j  \big),
\end{equation}
\begin{equation}
\begin{split}
\label{eq:dfeprope2} {\boldsymbol f}_{j,l} & = \beta_{j,l}
{\boldsymbol \Pi}_{j,l} \big( {\boldsymbol Q}^H \big( {\boldsymbol
R} - \beta_{j,l}{\boldsymbol Q} {\boldsymbol \Pi}_{j,l}
{\boldsymbol Q}^H \big)^{-1}  \cdot \\ & \quad \big({\boldsymbol
p}_{j} - \beta_{j,l}{\boldsymbol \Pi}_{j,l} {\boldsymbol t}_j
\big)
 - {\boldsymbol t}_{j}) .
\end{split}
\end{equation}
The above expressions only depend on statistical quantities, and
consequently on the channel matrix ${\boldsymbol H}$, the symbol
and noise variance $\sigma^2_s$ and $\sigma^2_n$, respectively,
and the constraints. However, the matrix inversion required for
computing ${\boldsymbol w}_{j,l}$ is different for each branch and
data stream, thereby rendering the scheme computationally less
efficient. The expressions obtained in (\ref{eq:dfeprop2}) and
(\ref{eq:dfeprop1}) are equivalent and only require iterations
between them for an equivalent performance.

Simplifying the equations in (\ref{eq:dfeprop1}) and
(\ref{eq:dfeprop2}), using the fact that the quantity
${\boldsymbol t}_j= {\boldsymbol 0}$ for interference
cancellation, ${\boldsymbol v}_{j,l}= {\boldsymbol 0}$, and
assuming perfect feedback (${\boldsymbol s} = \hat{\boldsymbol
s}$) we get
\begin{equation}
\label{eq:dfe1} {\boldsymbol w}_{j,l} = \big( {\boldsymbol
H}{\boldsymbol H}^H + {\sigma_n^2}/{\sigma_s^2 } {\boldsymbol I}
\big)^{-1} {\boldsymbol H} ({\boldsymbol \delta}_j + {\boldsymbol
f}_{j,l})
\end{equation}
\begin{equation}
\label{eq:dfe2} {\boldsymbol f}_{j,l} = \beta_{j,l} {\boldsymbol
\Pi}_{j,l} \big(\sigma_s^2{\boldsymbol H}^H {\boldsymbol w}_{j,l}
\big) ,
\end{equation}
where ${\boldsymbol \delta}_j = [ \underbrace{0 \ldots 0}_{j-1}
~1~ \underbrace{0 \ldots 0}_{N_T-j-2}]^T$ is a $N_T \times 1$
vector with a one in the $j$th element and zeros elsewhere. The
proposed MB-MMSE-DF detector expressions above require the channel
matrix ${\boldsymbol H}$ (in practice an estimate of it) and the
noise variance $\sigma_n^2$ at the receiver. In terms of
complexity, it requires for each branch $l$ the inversion of an
$N_R \times N_R$ matrix and other operations with complexity
$O(N_R^3)$. However, the expressions obtained in
(\ref{eq:dfeprop1}) and (\ref{eq:dfeprop2}) for the general case,
and in (\ref{eq:dfe1}) and (\ref{eq:dfe2}) for the case of perfect
feedback, reveal that the most expensive operations, i.e., the
matrix inversions, are identical for all branches. Therefore, the
design of filters for the multiple branches only requires further
additions and multiplications of the matrices. Moreover, we can
verify that the filters ${\boldsymbol w}_{j,l}$ and ${\boldsymbol
f}_{j,l}$ are dependent on one another, which means the designer
has to iterate them before applying the detector.

The MMSE associated with the pair of filters ${\boldsymbol
w}_{j,l}$ and ${\boldsymbol f}_{j,l}$ and the statistics of data
symbols $s_j[i]$ is given by
\begin{equation}
\begin{split}
{\rm MMSE} (s_j[i],{\boldsymbol w}_{j,l}, {\boldsymbol f}_{j,l}) &
= \sigma_s^2 - {\boldsymbol w}_{j,l}^H {\boldsymbol R}
{\boldsymbol
w}_{j,l} + {\boldsymbol f}_{j,l}^H {\boldsymbol f}_{j,l} 
\end{split}
\end{equation}
where $\sigma_s^2 = E[|s_j[i]|^2]$ is the variance of the desired
symbol.

\subsection{Design of Cancellation Patterns and Ordering}

We detail the design of the shape constraint matrices
${\boldsymbol S}_{j,l}$ and vectors ${\boldsymbol v}_{j,l}$,
motivate their choices and explain how the ordering of the data is
obtained. By pre-storing matrices for the $N_T$ data streams and
for the $L$ branches of the proposed MB-MMSE-DF detector, a
designer can exploit different patterns of cancellation and
orderings. 
Specifically, we are
interested in shaping the filters ${\boldsymbol f}_{j,l}$ for the
$N_T$ data streams and the $L$ branches with the matrices
${\boldsymbol S}_{j,l}$ such that resulting constraint vectors
${\boldsymbol v}_{j,l}$ are null vectors. This corresponds to
allowing feedback connections of only a subgroup of data streams.
For the first branch of detection ($l=1$), we can use the SIC
approach used in the VBLAST and
\begin{equation}
\begin{split}
{\boldsymbol S}_{j,l} {\boldsymbol f}_{j,l} & = {\boldsymbol 0},
~~
l=1 \\
{\boldsymbol S}_{j,l} & = \left[ \begin{array}{cc} {\boldsymbol
0}_{j-1} & {\boldsymbol 0}_{j-1, N_{T}-j+1} \\ {\boldsymbol
0}_{N_{T}-j+1,j-1} & {\boldsymbol I}_{N_{T}-j+1 } \end{array}
\right], ~~ j=1, \ldots, N_T,
\end{split}
\end{equation}
where ${\boldsymbol 0}_{m,n}$ denotes an $m \times n$-dimensional
matrix full of zeros, and ${\boldsymbol I}_m$ denotes an
$m$-dimensional identity matrix. For the remaining branches, we
adopt an approach based on permutations of the structure of the
matrices ${\boldsymbol S}_{j,l}$, which is given by
\begin{equation}
\begin{split}
{\boldsymbol S}_{j,l} {\boldsymbol f}_{j,l} & = {\boldsymbol 0},
~~
l=2, \ldots, L \\
{\boldsymbol S}_{j,l} & = \phi_l \left[ \begin{array}{cc}
{\boldsymbol 0}_{j-1} & {\boldsymbol 0}_{j-1, N_{T}-j+1} \\
{\boldsymbol 0}_{N_{T}-j+1,j-1} & {\boldsymbol I}_{N_{T}-j+1 }
\end{array} \right], ~~ j=1, \ldots, N_T,
\end{split}
\end{equation}
where the operator $\phi_l [ \cdot ]$ permutes the columns of the
argument matrix such that one can exploit different orderings via
SIC. These permutations are straightforward to implement and allow
the increase of the diversity order of the proposed MB-MMSE-DF
detector.

An alternative approach for shaping ${\boldsymbol S}_{j,l}$ for
one of the $L$ branches is to use a PIC approach and design the
matrices as follows
\begin{equation}
\begin{split}
{\boldsymbol S}_{j,l} {\boldsymbol f}_{j,l} & = {\boldsymbol 0},
~~
l  \\
{\boldsymbol S}_{j,l} & =  
{\rm diag}
~({\boldsymbol \delta}_j), ~~ j=1, \ldots, N_T, \label{picshape}
\end{split}
\end{equation}
The PIC requires the use of an initial vector of decisions taken
with the feedforward filters ${\boldsymbol w}_{j,l}$. The ordering
for the proposed MB-MMSE-DF detector is based on determining the
optimal ordering for the first branch, which employs a V-BLAST
type SIC, and then uses phase shifts for increasing the diversity
for the remaining branches. The proposed ordering for $l=1,
~\ldots,~L$ is given by
\begin{equation}
\begin{split}
\{ o_{1,l}, \ldots, o_{N_T,l} \} & = \arg \min_{ o_{1,l}, \ldots,
o_{N_T,l}} \sum_{l=1}^{L} \sum_{j=1}^{N_T} {\rm
MMSE} (s_j[i], {\boldsymbol w}_{j,l}, {\boldsymbol f}_{j,l}) 
\end{split}
\end{equation}
This algorithm finds the optimal ordering for each branch. For the
case of a single branch detector this corresponds to the optimal
ordering of the V-BLAST detector. The idea with the multiple
branches and their orderings is to attempt to benefit a given data
stream or group for each decoding branch. With this approach, a
data stream that for a given ordering appears to be in an
unfavorable position can benefit in other parallel branches by
being detected in a more favorable situation, increasing the
diversity of the proposed detector. 
%

\section{Multistage Detection for the MB-MMSE-DF}

In this section, we present a strategy based on iterative
multi-stage detection \cite{woodward2,delamare_spadf} that
gradually refines the decision vector, combats error propagation
and improves the overall performance. We incorporate this strategy
into the MB-MMSE-DF scheme and then investigate the improvements
to detection performance. An advantage of multistage detection
that has not been exploited for the design of MIMO detectors is
the possibility of equalizing the performance of the detectors
over the data streams. Since V-BLAST or DF detection usually
favors certain data streams (the last detected ones) with respect
to performance, it might be important for some applications to
yield uniform performance over the data streams. Specifically, the
MB-MMSE-DF detector with $M$ stages can be described by
\begin{equation}
{z}^{(m+1)}_{j,l}(i)= \tilde{\boldsymbol w}^{H}_{j,l}{\boldsymbol
r}[i] - \tilde{\boldsymbol f}^{H}_{j,l} \hat{\boldsymbol
s}^{o,(m)}_{j,l}[i], ~~ m=0,~1,~\ldots,~M,
\end{equation}
where the MMSE filters $\tilde{\boldsymbol w}_{j,l}$ and
$\tilde{\boldsymbol f}_{j,l}$ are designed with the approach
detailed in the previous subsection, $M$ denotes the number of
stages and $\hat{\boldsymbol s}^{o,(m)}_{j,l}[i]$ is the vector of
tentative decisions from the preceding iteration that is described
by:
\begin{equation}
\hat{ s}^{o,(1)}_{k,j,l}[i] = {\rm Q} \Big( {\boldsymbol
w}^{H}_{j,l}{\boldsymbol r}[i] \Big), ~~ k = 1,~\ldots, ~N_T,
\end{equation}
\begin{equation}
\hat{ s}^{o,(m)}_{k,j,l}[i] = {\rm Q} \Big( z^{(m)}_{j,l}[i]
\Big), ~ m = 2, ~\ldots, ~M,
\end{equation}
where the number of stages $M$ depends on the scenario.

In order to equalize the performance over the data streams
population, we consider an M-stage structure. The first stage is
an MB-MMSE-DF scheme with filters ${\boldsymbol w}_{j,l}$ and
${\boldsymbol f}_{j,l}$. The tentative decisions are passed to the
second stage, which consists of another MB-MMSE-DF scheme with
similar filters but use the decisions of the first stage and so
successively. The output of the second stage of the resulting
scheme is
\begin{equation}
z_{j,l}^{(m+1)}[i]=[{\boldsymbol T}{\boldsymbol w}_{j,l}]^{H
}{\boldsymbol r}[i] - [{\boldsymbol T}{\boldsymbol f}_{j,l}]^{H
}]\hat{\boldsymbol s}^{0, (m)}_{j,l}[i]
\end{equation}
where $z_{j,l}^{(m+1)}[i]$ is the output of $j$th data stream
after multistage detection with $M$ stages, ${\boldsymbol T}$ is a
square permutation matrix with ones along the reverse diagonal and
zeros elsewhere. When using multiple stages, it is beneficial to
demodulate the data streams successively and in reverse order
relative to the first branch of the previous MB-MMSE-DF detector.
The role of reversing the cancellation order in successive stages
is to equalize the performance of the users over the population or
at least reduce the performance disparities. 

\section{Simulations}

In this section, we assess the bit error rate~(BER) performance of
the proposed and analyzed MIMO detection schemes , namely, the
sphere decoder (SD), the linear \cite{duel_mimo}, the VBLAST (SIC)
\cite{vblast}, the S-DF \cite{dhahir} with MMSE estimators and the
proposed MB-MMSE-DF techniques without and with error propagation
mitigation techniques for the design of MIMO detectors. We also
consider the lattice-reduction aided versions of the linear and
the VBLAST detectors \cite{wuebben}, which are denoted
LR-MMSE-Linear and LR-MMSE-SIC, respectively. The channels'
coefficients are taken from complex Gaussian random variables with
zero mean and unit variance and QPSK modulation is employed. We
average the experiments over $100000$ runs, use packets with
$Q=200$ symbols and define the signal-to-noise ratio as
$\textrm{SNR} = 10 \log_{10} {N_T \sigma_s^2}/{\sigma^2_n}$, where
$\sigma_s^2$ is the variance of the symbols and $\sigma^2_n$ is
the noise variance.

\begin{figure}[!htb]
\begin{center}
\def\epsfsize#1#2{0.975\columnwidth}
\epsfbox{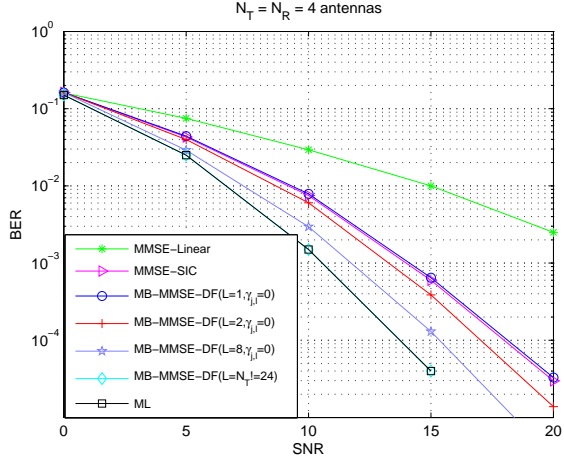} \vspace*{-1.5em} \caption{\footnotesize BER
Performance of the detectors with perfect channel estimation.}
\end{center}
\end{figure}

Let us first consider the proposed MB-MMSE-DF detector and
evaluate the number of branches $L$ that should be used for a MIMO
system with $N_T=N_R=4$ antennas with $\gamma_{j,l}=0$. We also
compare the proposed user ordering algorithm against the optimal
ordering approach, briefly described in Section 3, that tests
$N_T!=24$ possible branches and selects the most likely estimate.
We designed the MB-MMSE-DF detectors with $L=1,2 ~{\rm and}~8$
parallel branches, using one branch with the PIC of
(\ref{picshape}) for $L>1$, and compared their BER performance
against the SNR with the existing schemes. The results in Fig. 2
show that the MB-MMSE-DF detector outperforms the linear one by a
substantial margin and is comparable with the VBLAST for $L=1$.
The performance of the MB-MMSE-DF improves as the number of
parallel branches increases, resulting in improvements for more
than $L=1$ branches. For the case of $L=24$, we obtain a
performance identical to the optimal ML detector and for $L=8$, we
get a performance within $1.5$ dB from the optimal ML detector
computed with the SD.

\begin{figure}[!htb]
\begin{center}
\def\epsfsize#1#2{0.975\columnwidth}
\epsfbox{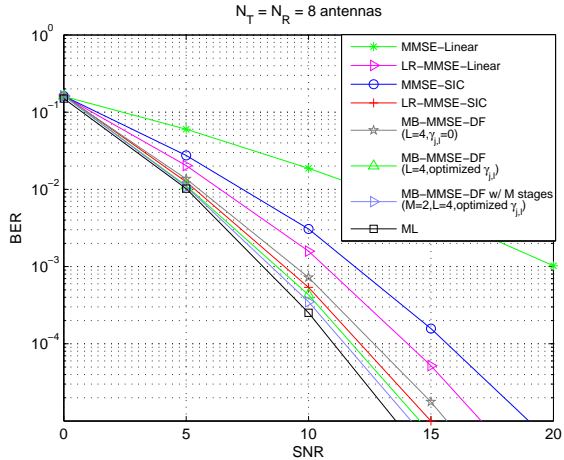} \vspace*{-1.5em} \caption{\footnotesize BER
Performance with perfect channel estimates.}
\end{center}
\end{figure}

In the next experiment, shown in Fig. 3, we compare the BER
performance of the proposed MB-MMSE-DF detector with $M=2$ stages,
$L=4$ ($3$ SICs and $1$ PIC shaping matrices) with perfect channel
estimation and error propagation mitigation. We include in
addition to the previous experiment the LR-MMSE-Linear and
LR-MMSE-SIC detection schemes \cite{wuebben} in the comparison.
The results depicted in Fig. 3 for a scenario with perfect channel
estimates shows that the proposed MB-MMSEDF detector achieves a
performance which is very close to the optimal ML implemented with
the SD and outperforms the linear, the VBLAST, the LR-MMSE-Linear
and LR-MMSE-SIC detectors by a significant margin.
\vspace*{-0.5em}

\section{Conclusions}

We proposed the MB-MMSE-DF detector for MIMO systems based on
multiple feedback branches. We also proposed the design of MMSE
filters subject to shape and magnitude constraints and the use of
multi-stage detection for error propagation mitigation. The
MB-MMSE-DF detector was compared with existing detectors and was
shown to approach the ML detector performance. Future work will
investigate low-complexity design of the filters, channel
estimation and the use for multiuser and multi-cell MIMO systems.

\end{document}